\documentclass[aip,pop,reprint]{revtex4-1}
\usepackage[utf8]{inputenc}
\usepackage[T1]{fontenc}
\usepackage{amsmath}
\usepackage{amsfonts}
\usepackage{amssymb}
\usepackage{graphicx}
\usepackage{epstopdf}
\usepackage{siunitx}
\usepackage{lmodern}
\usepackage[pdftex,colorlinks,linkcolor=blue,citecolor=blue,urlcolor=blue]{hyperref}
\newcommand{\abs}[1]{\left|#1\right|}

\newcommand{\ion}{\textup{i}}
\newcommand{\el}{\textup{e}}
\newcommand{\smax}{\textup{max}}
\newcommand{\smin}{\textup{min}}
\newcommand{\sth}{\textup{th}}
\newcommand{\subb}{\textup{b}}
\newcommand{\plasm}{\textup{p}}
\newcommand{\inj}{\textup{inj}}
\newcommand{\subf}{\textup{tr}}
\newcommand{\chan}{\textup{c}}

\begin{document}
\title{Beam loading in the bubble regime in plasmas with hollow channels}
\author{A.\,A.\,Golovanov}
\author{I.\,Yu.\,Kostyukov}
\affiliation{Institute of Applied Physics RAS, 603950 Nizhny Novgorod, Russia}
\author{J.\,Thomas}
\author{A.\,Pukhov}
\affiliation{Institut für Theoretische Physik I, Heinrich-Heine-Universität Düsseldorf, Düsseldorf D-40225, Germany}

\begin{abstract}
	Based on the already existing analytical theory of the strongly-nonlinear wakefield (which is called ``bubble'') in transversely inhomogeneous plasmas, we study particular behavior of non-loaded (empty) bubbles and bubbles with accelerated bunches.
	We obtain an analytical expression for the shape of a non-loaded bubble in a general case and verify it with particle-in-cell (PIC) simulations.
	We derive a method of calculation of the acceleration efficiency for arbitrary accelerated bunches.
	The influence of flat-top electron bunches on the shape of a bubble is studied.
	It is also shown that it is possible to achieve acceleration in a homogeneous longitudinal electric field by the adjustment of the longitudinal density profile of the accelerated electron bunch.
	The predictions of the model are verified by 3D PIC simulations and are in a good agreement with them.
\end{abstract}
\maketitle

\section{Introduction}
Lately a lot of attention has been given to plasma acceleration methods\cite{Esarey2009RevModPhys, Kostyukov2015UFN} in which a driver (an intense laser pulse \cite{Tajima1979laser} or a bunch of charged particles \cite{Rosenzweig1988experimental}) is used to excite a plasma wakefield whose large longitudinal field is used for acceleration.
Such methods provide acceleration rates orders of magnitude higher than conventional methods.
So far electron bunches with the energy up to \SI{4.2}{\GeV} at the acceleration distance of \SI{9}{\cm} have been observed in experiments for laser-wakefield acceleration (LWFA)\cite{Leemans2014PRL}, while for plasma-wakefield acceleration (PWFA) the possibility of energy doubling from \SI{42}{\GeV} to \SI{85}{\GeV} at a distance of approximately one meter has been demonstrated \cite{Blumenfeld2007Nature}.

One of the most promising regimes of plasma acceleration is the so-called ``bubble'' or ``blow-out'' regime in which electrons behind the driver are almost completely expelled and a spherical plasma cavity free of electrons is formed \cite{Pukhov2002Bubble}.
On its border a thin electron sheath consisting of the expelled electrons is created.
The cavity itself travels with a near-luminous velocity through the plasma.
The longitudinal electric field in it is uniform in the transversal direction, while the focusing force acting on the accelerated electrons is linear in radius and uniform along the cavity \cite{Kostyukov_2004_PoP_11_115256}.
In the current paper this cavity will be referred to as a ``bubble''.
In spite of the fact that this regime provides large acceleration gradients, obtaining electron bunches with low emittance, low energy spread, and high stability is a challenging task \cite{Malka2002electron}.
Such properties of electron bunches are important for applications, for example for their use in free electron lasers \cite{Jaroszynski2006radiation}.
Numerous methods dedicated to the improvement of the bunch quality have been proposed including self-guiding of the laser pulse \cite{Clayton_2010_PRL_105_105003}, guiding of the pulse in a preformed parabolic channel \cite{Leemans_2006_NP_2_100696, Leemans2014PRL}, different types of controlled injection \cite{Fubiani_2004_PRE_70_016402, Faure_2006_N_444_71200737, Chen_2006_JoAP_99_056109, Zeng_2015_PRL_114_084801}, etc.
One of such ways is the use of plasmas with deep (hollow) channels, which is commonly proposed for different types of plasma accelerators \cite{Chiou_1995_PoP_2_010310, Schroeder_1999_PRL_82_061177, Kimura_2011_PRSTAB_14_041301, Gessner_2016_Nat_Commun_7_11785}.
The lack of the ion column at the axis for such kind of channels significantly reduces the focusing force acting on the accelerated electrons, while, for the case of LWFA, the parameters of the channel also provide additional freedom for balancing between the laser depletion and dephasing lengths.
For instance, the possibility of obtaining electron bunches with the energy of \SI{7.5}{\GeV} and the energy spread of only 0.3\% in the bubble regime in a plasma with a deep channel has been shown in numerical simulations \cite{Pukhov2014Channel}.

Due to the complexity of the strongly nonlinear ``bubble'' regime of the plasma wakefield it is commonly studied using 3D particle-in-cell (PIC) simulations \cite{Pukhov_1999_JoPP_61_030425}.
However, its theoretical description is also of considerable interest.
Despite the nonlinear nature of the strongly non-linear wakefield, various approaches have led to the creation of phenomenological models of the bubble \cite{Kostyukov_2004_PoP_11_115256, Kostyukov_2009_PRL_103_175003}, as well as the development of a similarity theory \cite{Gordienko_2005_PoP_12_043109} and an analytical model describing the border of a bubble in homogeneous plasmas\cite{Lu2006nonlinear}.
Generalization of this theory to the case of plasmas with arbitrary radial inhomogeneity has recently been made in \citet{Thomas2015channels}. 
A subsequent generalization to arbitrary spatial distribution of electrons inside the sheath at the bubble's border has shown that the choice of this profile has little effect on the shape of the cavity \cite{Golovanov_2016_QE_46_295}.

The current paper uses the generalized theory to analytically study the blow-out regime for the case of plasmas with channels and to analyze beam loading for this case.
The results are obtained for arbitrary plasma profiles. 
Certain examples of density profiles, namely power-law ones and plasmas with deep channels, are used for illustrative purposes.
Power-law profiles are notable as many results for them can be obtained analytically, while plasmas with deep (vacuum) channels are of interest because, as previously mentioned, the transversal focusing force in them is almost completely suppressed \cite{Pukhov2014Channel}.
In section~\ref{sec:bubbleShape} we briefly describe the model introduced in \citet{Thomas2015channels} and simplify the equation for the boundary of the bubble derived in that paper to make further analysis easier.
In section~\ref{sec:nonloaded} we obtain the shape of a non-loaded bubble (i.e.\@ a bubble without accelerated particles) for an arbitrary plasma profile.
This solution is necessary for the further analysis of loaded bubbles.
The obtained shapes for the case of a plasma with a vacuum channel are compared to the results of 3D PIC simulations.
Section~\ref{sec:efficiency} is dedicated to the efficiency of the energy transfer from the electromagnetic fields in the bubble to accelerated bunches.
We propose the definition of the efficiency and a method of its calculation for arbitrary plasma profiles and arbitrary accelerated electron bunches.
Section~\ref{sec:flattop} then discusses the case of a flat-top accelerated bunch.
A threshold value of a charge density for which the efficiency reaches its maximum is calculated.
Finally, section~\ref{sec:constantField} focuses on the possibility of creating a homogeneous accelerating field by adjusting the profile of a witness electron bunch.
Such bunch profiles are found for arbitrary plasma density profiles.
The analytical results are also compared to the results of 3D PIC simulations.

\section{General equations}\label{sec:bubbleShape}

In our model we assume that we have boundless plasma in which a laser pulse or an electron driver propagates along the axis $z$ and excites a plasma wakefield in the bubble or blow-out regime.
The plasma is non-uniform in the plane perpendicular to the axis $z$ with the electron density $n_0(r)$ depending only on the distance $r$ from the axis.
Axial symmetry is assumed.
For simplicity we use dimensionless units in which all charges are normalized to $e$, densities to $n_\plasm$, time to $\omega_\plasm^{-1}$, coordinates to $k_\plasm^{-1} = c/\omega_\plasm$, momenta and energies to $mc$ and $mc^2$ respectively, and electric and magnetic fields to $mc\omega_\plasm/e$.
Here $e > 0$ is the elementary charge, $m$ is the electron mass, $c$ is the speed of light in vacuum, $n_\plasm$ is a typical plasma density (for example, for the case of a plasma with a hollow channel it may be the density outside the channel), $\omega_\plasm = (4\pi e n_\plasm/m)^{1/2}$ is a typical electron plasma frequency.

In analytical theories of the strongly-nonlinear regime a bubble is commonly described as a cavity completely free of electrons and surrounded by a thin electron sheath.
In this case the plasma outside the sheath is considered non-perturbed.
For example, such approach is used in \citet{Lu2006nonlinear}, where it has been shown that the bubble boundary (the inner border of the bubble sheath) can be described with a second-order ordinary differential equation (ODE).
As shown in \citet{Thomas2015channels}, the bubble boundary $r_\subb(\xi)$ for the case of radially-inhomogeneous plasma can be described with a similar equation
\begin{equation}
	A(r_\subb) \frac{d^2r_\subb}{d\xi^2} + B(r_\subb) \left(\frac{dr_\subb}{d\xi}\right)^2 + C(r_\subb) = \frac{\lambda(\xi)}{r_\subb}
	\label{bubbleShape:shapeEquation}
\end{equation}
where $\xi = t - z$ is the coordinate comoving with the bubble.
This equation is derived under the quasistatic approximation, thus all dependencies on time $t$ and the longitudinal coordinate $z$ are replaced with dependencies on their combination $\xi$.
Under certain conditions---which are discussed below---the coefficients of Eq.~(\ref{bubbleShape:shapeEquation}) are
\begin{equation}
	A = \frac{S_\ion(r_\subb)}{2},\quad
	B = \frac{\rho_\ion(r_\subb) r_\subb}{2},\quad
	C = \frac{S_\ion(r_\subb)}{2 r_\subb}.
	\label{bubbleShape:coefficients}
\end{equation}
These coefficients depend on the radial ion charge density profile $\rho_\ion(r)$ via the function 
\begin{equation}
	S_\ion(r) = \int_0^r \rho_\ion(r') r' dr'.
	\label{bubbleShape:Si}
\end{equation}
As opposed to a similar function with the opposite sign introduced in \citet{Thomas2015channels}, the function $S_\ion(r)$ is always positive, as $\rho_\ion > 0$.
The function $\lambda(\xi)$ in the right-hand side of Eq.~(\ref{bubbleShape:shapeEquation}) is determined by electron bunches inside the bubble:
\begin{equation}
	\lambda(\xi) = -\int_0^{r_\subb} J_z(\xi,r')r' dr'.
	\label{bubbleShape:lambda}
\end{equation}
Here $J_z(\xi, r)$ is the longitudinal current density created by electron bunches.
As typical velocities of accelerated and driving bunches are close to the speed of light $c$, this current density in the dimensionless units is approximately equal to the electron bunch charge density: $J_z(\xi, r) \approx \rho_\el(\xi, r)$.
Because $\rho_\el < 0$, the function $\lambda(\xi)$ is always positive for electron bunches which are considered in this paper.
However, for positron or proton bunches the sign is the opposite.

If the coefficients $A$, $B$, and $C$ are given by Eq.~(\ref{bubbleShape:coefficients}), the longitudinal electric field in the bubble is determined by its boundary $r_\subb(\xi)$ according to
\begin{equation}
	E_z(\xi) = \frac{S_\ion(r_\subb)}{r_\subb} \frac{dr_\subb}{d\xi}.
	\label{bubbleShape:Ez_initial}
\end{equation}
This electric field is accelerating for electrons situated in the rear part of the bubble ($dr_\subb/d\xi < 0$) and is decelerating in its front part where the driver is located.
It is also important that the longitudinal electric field retains the property of being uniform in the transverse direction even for plasmas with channels.

As already mentioned above, the simple expressions for the coefficients (\ref{bubbleShape:coefficients}) and for the longitudinal field (\ref{bubbleShape:Ez_initial}) are not universal and are valid only if the electron sheath is thin and the bubble is large enough.
To be more precise, the conditions $\Delta \ll r_\subb$ and $\Delta \gg 2 r_\subb / S_\ion(r_\subb)$ should be fulfilled\cite{Golovanov_2016_QE_46_295}.
Here $\Delta$ is the thickness of the electron sheath surrounding the bubble.
It is assumed to be independent of the longitudinal coordinate in our model.
Therefore, Eqs.~(\ref{bubbleShape:coefficients}), (\ref{bubbleShape:Ez_initial}) are not valid in the areas where the bubble local radius $r_\subb$ is small, i.e.\@ at the front and rear edges of the bubble.
Hence, these equations cannot be used to describe the process of the bubble excitation by an electron bunch.
However, this process will be out of scope of this paper because the evolution of the bubble boundary behind the driver does not depend on it.
This allows us not to consider the properties of the driver at all.
This fact also implies that the part of the bubble behind a laser driver is correctly described by Eq.~(\ref{bubbleShape:shapeEquation}) despite it containing no laser-related terms.

So, we assume that we have an arbitrary driver which excites a bubble with a maximum transverse size $R_\subb$.
For convenience we also assume that the maximum is reached at $\xi=0$, so the initial conditions of Eq.~(\ref{bubbleShape:shapeEquation}) are
\begin{equation}
	r_\subb(\xi = 0) = R_\subb,\quad \frac{dr_\subb}{d\xi}\bigg|_{\xi = 0} = 0.
	\label{bubbleShape:initialConditions}
\end{equation}
We also assume that the driver is fully located in the front part of the bubble (${\xi < 0}$).
Therefore, in the rear part of the bubble (${\xi > 0}$) the term $\lambda(\xi)$ is determined solely by witness electron bunches.

In its rear part the bubble collapses to the axis $r = 0$ due to the Coulomb attraction of the plasma ion column.
Hence, we will assume that $r_\subb(\xi)$ is monotonous for ${\xi > 0}$.
This monotonicity might be broken if the source $\lambda(\xi)$ in the right-hand side of Eq.~\eqref{bubbleShape:shapeEquation} is sufficiently large, i.e.\@ when an accelerated electron bunch has a charge large enough to prevent the collapse of the bubble.
If $r_\subb(\xi)$ is monotonous, its inverse function $\xi_\subb(r)$ exists.
Therefore, we can use $r$ as a new variable and obtain an equation for $Y(r) = r_\subb'(\xi_\subb(r))$ from Eq.~(\ref{bubbleShape:shapeEquation}):
\begin{equation}
	S_\ion(r) r Y \frac{dY}{dr} + \frac{dS_\ion(r)}{dr} r Y^2 + S_\ion(r) = 2 \Lambda(r),
\end{equation}
where $\Lambda(r)=\lambda(\xi_\subb(r))$.
Multiplying this equation by $S_\ion$ and dividing it by $r$ we get a total derivative in the left-hand side of the equation:
\begin{equation}
	\frac{d}{dr}\left(S_\ion^2 Y^2 \right) = \frac{2 S_\ion}{r} (2 \Lambda - S_\ion).
\end{equation}
Integrating this equation with the initial conditions (\ref{bubbleShape:initialConditions}) and returning to the function $r_\subb(\xi)$, we obtain the equation
\begin{align}
	&\frac{dr_\subb(\xi)}{d\xi} = -\frac{1}{S_\ion(r_\subb)} \sqrt{ 2 F_\ion(r_\subb, R_\subb) -\int_{r_\subb}^{R_\subb} \frac{4 S_\ion \Lambda}{r'}dr'},
	\label{bubbleShape:rbPrime}\\
	&F_\ion(r_1, r_2) = \int_{r_1}^{r_2} \frac{S_\ion^2(r')}{r'} dr'.
\end{align}
The negative sign before the square root is chosen based on the fact that $dr_\subb/d\xi < 0$ in the rear part of the bubble.
As the function $S_\ion(r)$, according to its definition (\ref{bubbleShape:Si}), is at least quadratic in the neighborhood of $r=0$, dividing it by $r$ in Eq.~(\ref{bubbleShape:rbPrime}) does not lead to any singularities.
The longitudinal electric field, according to Eq.~(\ref{bubbleShape:Ez_initial}), is
\begin{equation}
	E_z(\xi) = - \frac{1}{r_\subb} \sqrt{2 F_\ion(r_\subb, R_\subb) - \int_{r_\subb}^{R_\subb} \frac{4 S_\ion \Lambda}{r'}dr'}.
	\label{bubbleShape:Ez}
\end{equation}
A similar approach to the derivation of the first-oder equation has been used in \citet{TzoufrasPhysPlasmas2009Beam} for homogeneous plasmas.

In the most general case Eq.~(\ref{bubbleShape:rbPrime}) is not an ODE because the function $\Lambda(r)$ depends on its solution and therefore is also unknown.
However, in several special cases this equation can be drastically simplified and solved analytically.
These cases are discussed next.

\section{Non-loaded bubble} \label{sec:nonloaded}
Let us consider a non-loaded bubble, i.e.\@ a bubble without accelerated electron bunches.
In this case $\lambda(\xi) = 0$, and Eq.~(\ref{bubbleShape:rbPrime}) can easily be integrated leading to the bubble boundary defined by
\begin{equation}
	\xi_\subb(r) = \int_{r}^{R_\subb} \frac{S_\ion(r') dr'}{\sqrt{2 F_\ion(r', R_\subb)}}.
	\label{nonloaded:shape}
\end{equation}
This equation can be used to find the length of the rear part of the bubble by assuming $r = 0$.
For specific plasma profiles the boundary of the bubble can be described by special functions.
In particular, it is done in \citet{TzoufrasPhysPlasmas2009Beam} for homogeneous plasmas and in \citet{Thomas2015channels} for plasmas with power-law channels (i.e.\@ channels with the power-law plasma density profile).

According to Eq.~(\ref{bubbleShape:Ez}), the longitudinal electric field in a non-loaded bubble is
\begin{equation}
	E_z(\xi) = - \frac{1}{r_\subb(\xi)}\sqrt{2 F_\ion(r_\subb, R_\subb)}.
	\label{nonloaded:Ez}
\end{equation}
Due to the fact that the function $r_\subb(\xi)$ is monotonous, $E_z(\xi)$ also has to be monotonous.
Therefore, it is impossible to find a plasma profile for which the longitudinal field in a non-loaded bubble is homogeneous (i.e.\@ does not depend on $\xi$).

From general formulas (\ref{nonloaded:shape}) and (\ref{nonloaded:Ez}) it is also possible to find the longitudinal field in the central part of the bubble, i.e.\@ near $\xi = 0$.
The boundary in the neighborhood of $\xi = 0$ is
\begin{equation}
	r_\subb(\xi) \approx R_\subb - \frac{\xi^2}{2 R_\subb},
	\label{nonloaded:shapeAtCenter}
\end{equation}
which corresponds to a sphere of a radius $R_\subb$.
The longitudinal field is zero at $\xi = 0$, therefore near the center of the bubble it can be approximated with a linear function
\begin{equation}
	E_z\approx - \xi \frac{S_\ion(R_\subb)}{R_\subb^2}.
	\label{nonloaded:EzAtCenter}
\end{equation} 
If we consider a channel in plasma, so that $\rho_\ion(r) \leq 1$, the function $S_\ion$ is limited ($S_\ion(r) \leq r^2/2$) and therefore $\abs{E_z} \leq \xi/2$.
So, the gradient of the electric field in the center of a bubble in a plasma channel is limited by the value $1/2$ which is reached for the case of homogeneous plasma.

The influence of power-law channels on the shape of the bubble has been already discussed in \citet{Thomas2015channels}.
It has been shown that the increase of the channel width and depth leads to the contraction of the bubble.
Here we check if similar behavior is observed for the case of plasma with a vacuum channel for which the ion density is modeled as $\rho_\ion = \theta(r_\chan-r)$, where $r_\chan$ is the radius of the channel and $\theta(x)$ is the Heaviside step function.
For this profile the function $S_\ion(r)$ defined by (\ref{bubbleShape:Si}) as well as all the integrals including this function can be analytically calculated.
Therefore, $\xi_\subb(r)$ determined by Eq.~(\ref{nonloaded:shape}) is calculated using a single integral instead of a double integral.

\begin{figure*}[tb]
	\centering
	\includegraphics{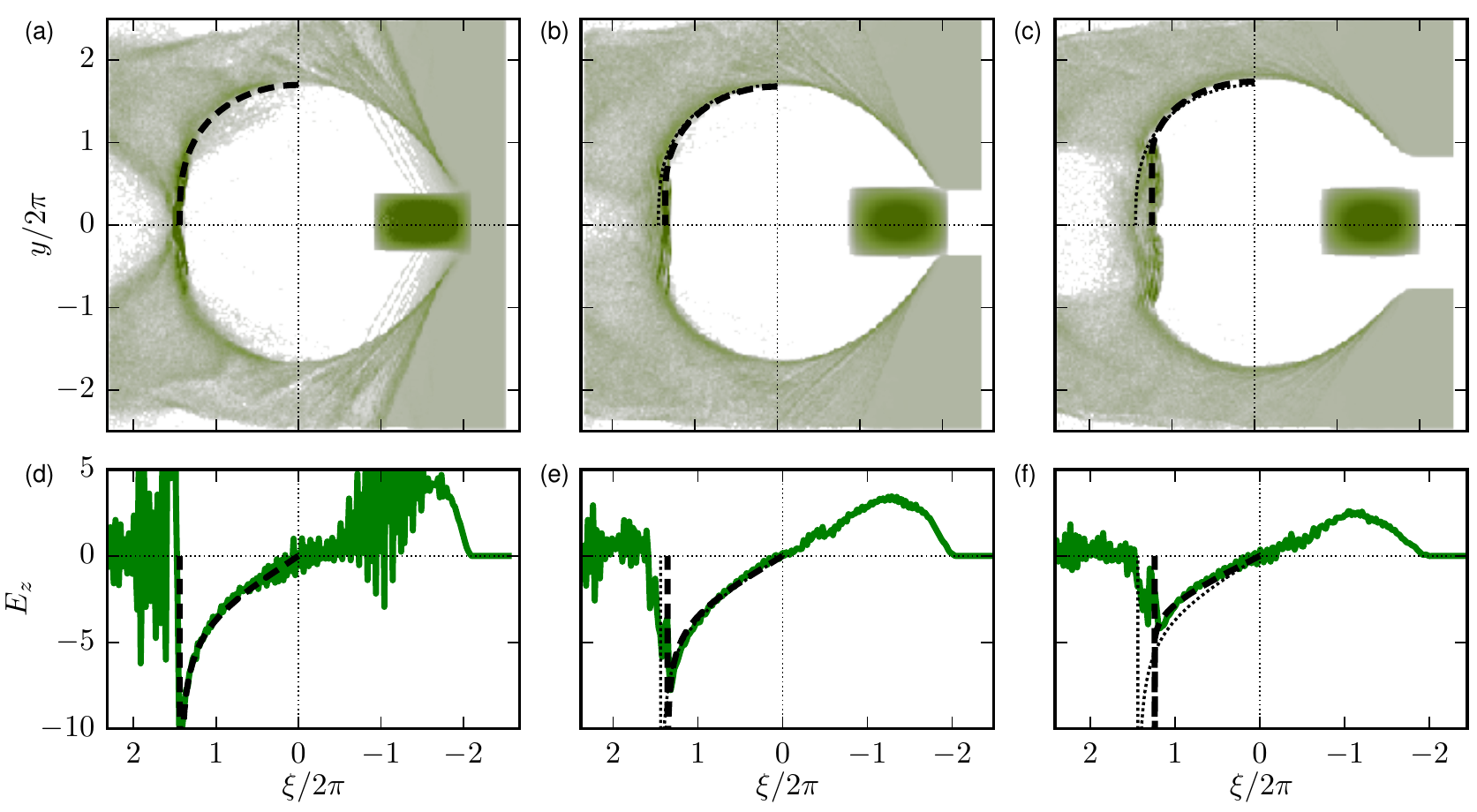}
	\caption{Electron density distributions (a--c) and corresponding longitudinal electric field profiles on the axis of the bubbles (d--f) as observed in PIC simulations for different channel radii $r_\chan = 0\text{, }0.4\pi\text{, and }0.8\pi$, respectively.
	Analytical solutions for the bubbles' boundaries and the fields in them calculated using Eqs.~(\ref{nonloaded:shape}) and (\ref{nonloaded:Ez}), respectively, are shown with the dashed lines.
	The dotted lines in (b, c, e, f) also show the solution for $r_\chan = 0$ for comparison.
	All lengths are normalized to $\lambda_\plasm = \SI{5}{\um}$.}
	\label{fig:nonloaded:pic}
\end{figure*}

The resulting boundaries of non-loaded bubbles for different values of $r_\chan$ and the corresponding longitudinal fields are shown in Fig.~\ref{fig:nonloaded:pic} in comparison to the results of PIC simulations carried out using the code \textsc{Quill}.
In these simulations we used plasma with a density outside the channel $n_\plasm = \SI{4.5e19}{\cm^{-3}}$ corresponding to a plasma wavelength of $\lambda_\plasm = \SI{5}{\um}$.
In all three cases an electron bunch with a total charge of \SI{3.2}{\nano\coulomb}, particle energy of \SI{1}{\GeV}, and typical longitudinal and transverse sizes of \SI{6}{\um} and \SI{4}{\um}, respectively, was used as a driver.
From Fig.~\ref{fig:nonloaded:pic} it is obvious that the length of the bubble and the gradient of the field in it decreases with the increase of a channel radius.
The figure also shows good correspondence between the results of the simulations and the predictions of the model both for the bubble boundaries and the longitudinal electric fields in them.

\section{Acceleration efficiency}\label{sec:efficiency}
As we have found the solution for a non-loaded bubble, we may now consider a bubble with an accelerated electron bunch.
One of the key points of efficient acceleration is the usage of the largest possible percentage of the energy pumped into the bubble by its driver.
Hence, the aim of this section is to describe the efficiency of the energy transfer from the bubble to the accelerated bunch.

Let us assume that a cylindrically symmetric electron bunch with an arbitrary profile $\lambda(\xi)$ is placed inside a bubble of the size $R_\subb$, and that its front and trailing edges have the coordinates $\xi_\inj$ and $\xi_\subf$, respectively.
The boundary of this bubble is described with the function $r_\subb(\xi)$.
Let us assume that the bubble boundary reaches zero at $\xi_\smax > \xi_\subf$, i.e.\@ the whole electron bunch is situated inside the bubble.
In this case the points $\xi_\inj$ and $\xi_\subf$ correspond to certain transverse sizes of the bubble $r_{\inj,\subf} = r_\subb(\xi_{\inj,\subf})$.
Consequently, Eq.~(\ref{bubbleShape:rbPrime}) for the bubble shape for $r_\subb<r_\subf$ becomes
\begin{equation}
	\frac{dr_\subb(\xi)}{d\xi} = -\frac{1}{S_\ion} \sqrt{2F_\ion(r_\subb, R_\subb) - \int_{r_\subf}^{r_\inj} \frac{4 S_\ion \Lambda}{r'}dr'}.
\end{equation}
One may notice that this equation corresponds to the equation for a non-loaded bubble with a different maximum radius $R'_\subb$ determined by
\begin{equation}
	F_\ion(0, R'_\subb) = F_\ion(0, R_\subb) - \int_{r_\subf}^{r_\inj} \frac{2 S_\ion(r') \Lambda(r')}{r'}dr'.
\end{equation}
As $R'_\subb$ has to be positive (otherwise the radicand would always be negative), we obtain limitations on $\Lambda(r)$:
\begin{equation}
	\int_{r_\subf}^{r_\inj} \frac{2 S_\ion(r') \Lambda(r')}{r'}dr' \leq F_\ion(0, R_\subb).
	\label{efficiency:inequationLambda}
\end{equation}
The case when this inequality does not hold corresponds to electron bunches in which electrons in the trailing edge of the bunch are decelerated instead of being accelerated (an example is given in Sec.~\ref{sec:flattop}).
We will not consider this case here and will assume that all electrons in the bunch are accelerated.
Then we can introduce a parameter characterizing the efficiency of the energy transfer from the bubble to the bunch:
\begin{equation}
	\eta = \frac{1}{F_\ion(0, R_\subb)} \int_{r_\subf}^{r_\inj} \frac{2 S_\ion(r') \Lambda(r')}{r'}dr'.
	\label{efficiency:efficiency_vs_Lambda}
\end{equation}
Due to Eq.~\eqref{efficiency:inequationLambda} this parameter can take values from 0 to 1.
In order to understand its physical meaning we can make a transformation $r' = r_\subb(\xi')$ and, by using Eqs.~(\ref{bubbleShape:lambda}) and (\ref{bubbleShape:Ez_initial}), obtain
\begin{multline}
\int_{r_\subf}^{r_\inj} \frac{2 S_\ion \Lambda}{r'}dr' \\ 
= 2 \int_{\xi_\inj}^{\xi_\subf}  \int_0^{r_\el} E_z(\xi') J_z(\xi', r') dr'd\xi' = \frac{1}{\pi} P,
\end{multline}
where $P$ is the total power consumed by the accelerated electron bunch as a result of the accelerating force action.
Therefore, the introduced efficiency (\ref{efficiency:efficiency_vs_Lambda}) measures the ratio between the power consumed by the accelerated bunch and the maximum possible power for the specified bubble.
Alternatively it may be written as
\begin{equation}
\eta = \frac{P}{P_\smax} \\ 
= - \frac{2}{F_\ion(0, R_\subb)} \int_{\xi_\inj}^{\xi_\subf} E_z(\xi') \lambda(\xi') d\xi'.
\label{theoryConsequences:efficiency_vs_lambda}
\end{equation}
This expression makes it possible to calculate the acceleration efficiency for any electron bunch.

\section{Bubble with a flat-top bunch} \label{sec:flattop}
Let us consider a bubble with the simplest electron bunch profile, i.e.\@ an infinitely long flat-top electron bunch, whose front edge has the longitudinal coordinate $\xi_\inj$.
The source $\lambda(\xi)$ for this type of the bunch is $\lambda(\xi) = \lambda_0 \theta(\xi - \xi_\inj)$, where $\theta(x)$ is the Heaviside step function.
For $\xi < \xi_\inj$ the source is zero, therefore the solution (\ref{nonloaded:shape}) for a non-loaded bubble remains valid.
From this solution we can obtain the transversal size of the bubble $r_\inj = r(\xi_\inj)$ for the coordinate corresponding to the front edge of the bunch.
Then the function $\Lambda(r)$ in Eq.~(\ref{bubbleShape:rbPrime}) is $\Lambda(r) = \lambda_0 \theta(r_\inj - r)$, therefore for $r < r_\inj$ this equation becomes
\begin{equation}
	\frac{dr_\subb}{d\xi} = -\frac{1}{S_\ion} \sqrt{2 F_\ion(r_\subb, R_\subb) - \lambda_0 \int_{r_\subb}^{r_\inj} \frac{4S_\ion(r')}{r'}dr'}.
	\label{flattop:rbPrime}
\end{equation}
For any point $r_\subb$ there is a threshold value of $\lambda_0$ for which the radicand is zero at this point:
\begin{equation}
	\hat\lambda_\sth(r_\subb) = F_\ion(r_\subb, R_\subb) \left( \int_{r_\subb}^{r_\inj} \frac{2S_\ion(r')}{r'}dr'\right)^{-1}.
\end{equation}
It can be shown that $\hat\lambda_\sth$ is a monotonically increasing function of $r_\subb$, therefore there is a global threshold value $\lambda_\sth$ defined by
\begin{equation}
	\lambda_\sth = \hat\lambda_\sth(0) = F_\ion(0, R_\subb) \left( \int_0^{r_\inj} \frac{2S_\ion(r')}{r'}dr' \right)^{-1}.
	\label{flattop:lambdath}
\end{equation}

If $\lambda_0 < \lambda_\sth$, the derivative $dr_\subb/d\xi$ is always negative, which means that $r_\subb(\xi)$ is a monotonous function and the bubble boundary reaches the axis $r = 0$.
If the value of $\lambda_0$ exceeds the threshold value, there is a certain point $r_0$ where the radicand and, consequently, $dr_\subb/d\xi$ reach zero, which means that the charge density of the electron bunch becomes sufficiently large to prevent the collapse of the rear part of the bubble; instead, the bubble expands after reaching this point.
Due to the fact that the sign of the longitudinal electric field inside the bubble is determined by the sign of $dr_\subb/d\xi$, the field changes from accelerating to decelerating beyond this point.
Therefore, it is sensible to place the trailing edge of a flat-top accelerated bunch at the point where $dr_\subb/d\xi = 0$ for this bunch.
As a result, the length of a flat-top bunch is limited either by the length of the resulting bubble (if the charge density is less then the threshold value) or by the point where the accelerating field reaches zero and would change to decelerating if we made the bunch any longer.

Knowing these limitations on the flat-top bunch's length, we can calculate the efficiency which can be reached with the use of a flat-top bunch.
To do so we use Eq.~(\ref{efficiency:efficiency_vs_Lambda}).
If $\lambda_0 < \lambda_\sth$, the bubble boundary reaches $r = 0$, therefore $r_\subf = 0$ and the expression for the efficiency is
\begin{equation}
	\eta = \frac{\lambda_0}{F_\ion(0, R_\subb)} \int_{0}^{r_\inj} \frac{2 S_\ion(r')}{r'}dr' = \frac{\lambda_0}{\lambda_\sth}.
	\label{flattop:efficiency_1}
\end{equation}
This expression shows that the efficiency grows linearly from $0$ to $1$ with the increase of $\lambda_0$ until the charge density reaches the threshold value.

For $\lambda_0 > \lambda_\sth$ there is a minimum value $r_0$ of the boundary's radial coordinate, therefore we should set $r_\subf = r_0$.
The value of $r_0$ can be found from Eq.~(\ref{flattop:rbPrime}) using the fact that $dr_\subb/d\xi = 0$ for $r_\subb = r_0$.
The resulting equation for the efficiency is
\begin{equation}
	\eta = \frac{F_\ion(r_0, R_b)}{F_\ion(0, R_b)}.
	\label{flattop:efficiency_2}
\end{equation}
With the increase of $\lambda_0$ the value of $r_0$ also increases and therefore the efficiency decreases.

Thus, the efficiency of the energy transfer from the bubble to the accelerated bunch reaches its maximal value of 1 for a flat-top bunch with the threshold charge density (i.e.\@ $\lambda_0 = \lambda_\sth$) for the specified injection point $\xi_\inj$.
For charge densities larger or smaller than the threshold value the efficiency drops below 1.
If we use a bunch shorter then optimal for the specified $\lambda_0$ and $\xi_\inj$, then the efficiency becomes even less than the one given by Eqs.~(\ref{flattop:efficiency_1}), (\ref{flattop:efficiency_2}).

\begin{figure}[tb]
	\centering
	\includegraphics{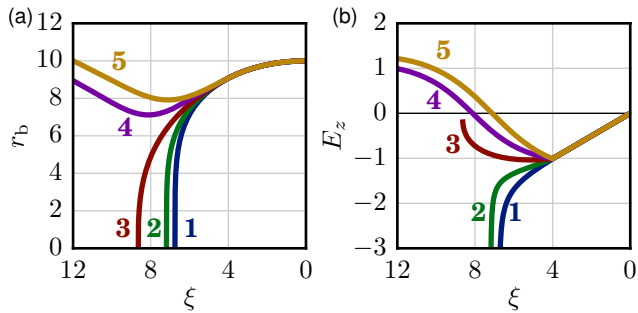}
	\caption{Dependencies of (a)~the bubble boundaries $r_\subb$ and (b)~the longitudinal fields $E_z$ on the longitudinal coordinate $\xi$ for different electron bunch charge densities $\lambda_0$.
	The dependencies are calculated numerically using Eqs.~(\ref{bubbleShape:shapeEquation}) and (\ref{bubbleShape:Ez_initial}), respectively. 
	Lines 1--5 show the solutions for the densities $\lambda_0$ from $0$ to $2\lambda_\sth$ in $0.5\lambda_\sth$ increments.
	The plasma profile used for calculations is parabolic ($n=2$), the maximum radius of the bubbles $R_\subb = 10$, the electron bunch injection coordinate $\xi_\inj = 4$.}
	\label{fig:flattop:powerlaw_shape}
\end{figure}

Let us for example consider the case of a power-law plasma profile $\rho_\ion = (r / R_\subb)^n$.
The normalization of density is chosen in the way that density is equal to 1 at the point where the bubble reaches its maximum transverse size.
Typical boundaries of bubbles with flat-top electron bunches and corresponding electric fields are shown in Fig.~\ref{fig:flattop:powerlaw_shape}.
This picture shows that the presence of an accelerated bunch leads to the elongation of the bubble compared to the case of a non-loaded bubble (line~1 in Fig.~\ref{fig:flattop:powerlaw_shape}).
The picture also confirms that there is a predicted threshold value $\lambda_\sth$ (line~3) which separates bubbles which collapse to $r=0$ from bubbles which continue expanding.

The threshold value determined by Eq.~(\ref{flattop:lambdath}) is
\begin{equation}
	\lambda_\sth = \frac{R_\subb^{n+4}}{4(n+2)r_\inj^{n+2}}.
\end{equation}
This value grows with the increase of the injection coordinate $\xi_\inj$ (larger $\xi_\inj$ corresponds to smaller $r_\inj$).
Therefore, the minimum threshold value is reached for $\xi_\inj = 0$ (and thus $r_\inj = R_\subb$) and is equal to
\begin{equation}
	\lambda_\smin = \min \lambda_\sth = \frac{R_\subb^2}{4(n+2)}.
\end{equation}
This minimum value decreases with the increase of the plasma profile exponent $n$.
It is also proportional to the square of the bubble size.

\begin{figure}[tb]
	\centering
	\includegraphics{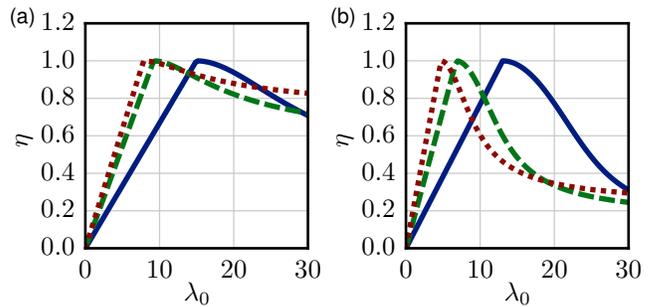}
	\caption{Dependencies of the efficiency $\eta$ on the charge density $\lambda_0$ calculated using Eq.~(\ref{flattop:powerlaw_efficiency}) for different exponents $n$ ($n = 0$, 2, 4 for the solid, dashed, and dotted lines, respectively) for the injection point (a) $\xi_\inj = 2$ and (b) $\xi_\inj = 4$.
	The maximum radius of the bubble $R_\subb = 10$.}
	\label{fig:flattop:powerlaw_efficiency}
\end{figure}

The efficiency for this kind of profile can be calculated analytically from Eqs.~(\ref{flattop:efficiency_1}) and (\ref{flattop:efficiency_2}) and is
\begin{equation}
	\begin{aligned}
		&\eta = \frac{\lambda_0}{\lambda_\sth},&&\lambda_0<\lambda_\sth, \\
		&\eta = 1 - X^2, &&\lambda_0>\lambda_\sth,
	\end{aligned}
	\label{flattop:powerlaw_efficiency}
\end{equation}
where
\begin{equation}
	X = \frac{\lambda_0}{2\lambda_\smin} - \sqrt{\left(\frac{\lambda_0}{2 \lambda_\smin}\right)^2 - \frac{\lambda_0}{\lambda_\sth} + 1}.
\end{equation}
Typical dependencies given by Eq.~(\ref{flattop:powerlaw_efficiency}) are shown in Fig.~\ref{fig:flattop:powerlaw_efficiency} for different plasma profiles and different injection points. 
It is visible that the threshold value of the charge density $\lambda_\sth$ is lower for plasmas with deeper channels corresponding to larger values of $n$.
One may also notice that the efficiency drops more rapidly with the increase of $\lambda_0$ beyond the threshold for a smaller injection point $\xi_\inj$.
It can be explained by the fact that for smaller $\xi_\inj$ the transverse velocities of the electrons in the bubble sheath at this point are smaller, and thus the electrons in the sheath are more sensitive to the variations of the charge density of the witness bunch.

\section{Uniform accelerating field by the adjustment of the bunch profile}
\label{sec:constantField}
In order not to increase the energy spread of the electrons in the accelerated bunch it is important to accelerate them in a uniform longitudinal field $E_z$.
This field is always uniform in the transverse direction in the scope of our model, therefore we are interested in reaching only the longitudinal homogeneity.
As shown in section~\ref{sec:nonloaded}, it is impossible to create a homogeneous longitudinal field in a non-loaded bubble.
However, the accelerated bunch's charge influences the shape of the bubble, which may be used to achieve the uniform field.
For homogeneous plasma it is done in \citet{TzoufrasPhysPlasmas2009Beam}.

Let us assume that the front edge of the electron bunch has the coordinate $\xi_\inj > 0$ which is referred to as the injection point.
Then for $0 \leq \xi \leq \xi_\inj$ there are no sources in the right-hand side of  Eq.~(\ref{bubbleShape:shapeEquation}) and the solution (\ref{nonloaded:shape}) for a non-loaded bubble is valid, which gives us the transversal size of the bubble $r_\inj = r(\xi_\inj)$ at the injection point.
The longitudinal field at this point can be found from Eq.~(\ref{nonloaded:Ez}) and is
\begin{equation}
	E_\inj = - E_z(r_\inj) = \frac{1}{r_\inj} \sqrt{2 F_\ion(r_\inj, R_\subb)}.
	\label{constantField:Et}
\end{equation}
The positive sign of $E_\inj$ is chosen for convenience.

For $\xi > \xi_\inj$ we try to find such a profile $\lambda(\xi)$ that the longitudinal field remains constant at the level of $-E_\inj$.
Using Eq.~(\ref{bubbleShape:Ez}) to find the required function $\Lambda(r_\subb)$ and Eq.~(\ref{bubbleShape:Ez_initial}) to find the bubble boundary $r_\subb(\xi)$, we obtain $\lambda(\xi) = \Lambda(r_\subb(\xi))$ in a parametric form
\begin{align}
	&\Lambda(r_\subb) = \frac{S_\ion(r_\subb)}{2} + \frac{E_\inj^2 r_\subb^2}{2 S_\ion(r_\subb)},
	\label{constantField:Lambda}\\
	&E_\inj (\xi - \xi_\inj) = \int_{r_\subb}^{r_\inj} \frac{S_\ion(r')}{r'} dr'.
	\label{constantField:xi_vs_rb}
\end{align}
Thus, it has been shown that the electron bunch profile $\lambda(\xi)$ providing homogeneous accelerating field can be found for any plasma profile $\rho_\ion(r)$.

As the length of the accelerated bunch is limited by the length of the bubble, we can find the maximum possible length of the accelerated bunch for any injection point $\xi_\inj$ using Eq.~(\ref{constantField:xi_vs_rb}) and setting $r_\subb = 0$:
\begin{equation}
\Delta \xi = \frac{1}{E_\inj} \int_0^{r_\inj} \frac{S_\ion}{r'} dr'.
\end{equation}
Knowing the maximum length of the bunch and the charge density $\lambda(\xi)$ in it, we can also calculate the maximum total charge of the bunch:
\begin{equation}
Q_\smax = 2\pi \int_{\xi_\inj}^{\xi_\inj+\Delta \xi} \lambda(\xi') d\xi' = \frac{2\pi F_\ion(0, R_\subb)}{E_\inj}.
\label{constantField:Qmax}
\end{equation}
It is worth mentioning that total power consumed by such a bunch and defined by $P = Q_\smax E_\inj$ does not depend on the value of the field $E_\inj$.
Therefore, by choosing the injection point $\xi_\inj$ we can balance between a high acceleration rate or a high value of the accelerated charge.

It is also worth mentioning that the average charge density $\langle \lambda \rangle$ of the bunch with the maximum length is equal to the threshold charge density $\lambda_\sth$ defined by Eq.~(\ref{flattop:lambdath}) for the case of a flat-top bunch.
At the same time, the efficiency of the energy transfer (\ref{efficiency:efficiency_vs_Lambda}) for such bunch is also equal to 1.

So, the electron bunch profile providing homogeneous longitudinal electric field in the whole volume of the bunch and leading to the acceleration with the efficiency up to $100\%$ has been found for an arbitrary plasma density profile.

\subsection{Power-law plasma profile}

As the first example let us consider plasmas with the power-law density profile $\rho_\ion = (r / R_\subb) ^ n$.
For these plasmas the function $\lambda(\xi)$ determined parametrically by Eqs.~(\ref{constantField:Lambda}) and (\ref{constantField:xi_vs_rb}) can be found analytically and is
\begin{multline}
	\lambda(\xi) = \frac{r_\inj^{n+2}}{2(n+2)R_\subb^n} - \frac{(n+2) E_\inj}{2}(\xi - \xi_\inj) \\
	 + \frac{R_\subb^n E_\inj^2 (n+2)}{2} \frac{1}{r_\subb^n(\xi)},
	\label{constantField:powerlaw_lambda}
\end{multline}
where
\begin{equation}
	r_\subb(\xi) = \left[r_\inj^{n+2} - R_\subb^n E_\inj (\xi - \xi_\inj) (n+2)^2 \right]^{\frac{1}{n+2}}
\end{equation}
and $E_\inj$ can be found from Eq.~(\ref{constantField:Et}).

\begin{figure}[tb]
	\includegraphics{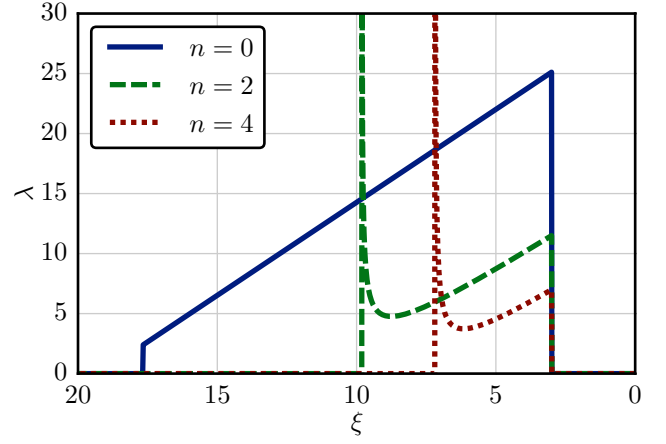}
	\caption{Dependencies of the charge density $\lambda$ in the accelerated bunches on the longitudinal coordinate $\xi$ calculated using Eq.~(\ref{constantField:powerlaw_lambda}) for different exponents $n$ of the power-law density profiles.
	The injection coordinate $\xi_\inj = 3$, the bubble maximum radius $R_\subb = 10$.}
	\label{fig:constantField:powerlaw_profiles}
\end{figure}

The bunch profiles determined by Eq.~(\ref{constantField:powerlaw_lambda}) for plasma profiles with different exponents $n$ are shown in Fig.~\ref{fig:constantField:powerlaw_profiles}.
This picture shows that the dependence of the charge density $\lambda$ on the longitudinal coordinate $\xi$ for such profiles is almost linear.
In the case of homogeneous plasma ($n = 0$) the bunch profile is strictly trapezoidal, while for $n > 0$ there is a singularity at the trailing edge of the bunch caused by the fact that the plasma density reaches $0$ at $r = 0$.
This singularity makes it difficult to use the electron bunch with the maximum length, thus further limiting the efficiency of the energy transfer.
It is also visible that the total charge in the accelerated bunch is smaller for deeper channels, i.e.\@ larger values of $n$.

\begin{figure}[tb]
	\centering
	\includegraphics{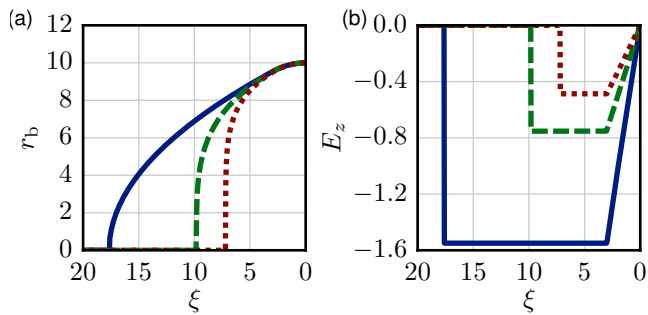}
	\caption{Dependencies of (a)~the bubbles boundaries $r_\subb$ and (b)~the longitudinal electric fields $E_z$ in them on the longitudinal coordinate $\xi$ calculated numerically using Eqs.~(\ref{bubbleShape:shapeEquation}) and (\ref{bubbleShape:Ez_initial}), respectively, for different exponents $n$ and electron bunches determined by Eq.~(\ref{constantField:powerlaw_lambda}).
	The solid, dashed, and dotted lines correspond to $n = 0$, $2$, $4$, respectively.
	The maximum bubble radius $R_\subb = 10$, the injection coordinate $\xi_\inj = 3$.}
	\label{fig:constantField:powerlaw_shape}
\end{figure}

Fig.~\ref{fig:constantField:powerlaw_shape} shows the bubbles corresponding to the profiles in Fig.~\ref{fig:constantField:powerlaw_profiles} and the longitudinal electric fields in them.
This image demonstrates that the longitudinal field is indeed homogeneous in the region where the bunch is present.
It is also seen that the presence of an electron bunch significantly increases the length of the bubble.
Figs.~\ref{fig:constantField:powerlaw_profiles} and \ref{fig:constantField:powerlaw_shape} also demonstrate the fact that the total acceleration power is much lower for channels with deeper channels (i.e.\@ with higher values of $n$), which is also supported by Eq.~(\ref{constantField:Qmax}) which for the power-law profiles is written as
\begin{equation}
	Q_\smax = \frac{1}{E_\inj} \frac{\pi R_\subb^2}{(n+2)^3}.
\end{equation}
This equation shows that thincreasee product $Q_\smax E_\inj$ equal to the acceleration power rapidly decreases with the increase of $n$ corresponding to the deepening and widening of the channel.

\subsection{Plasma with a vacuum channel}

\begin{figure}[tb]
	\centering
	\includegraphics{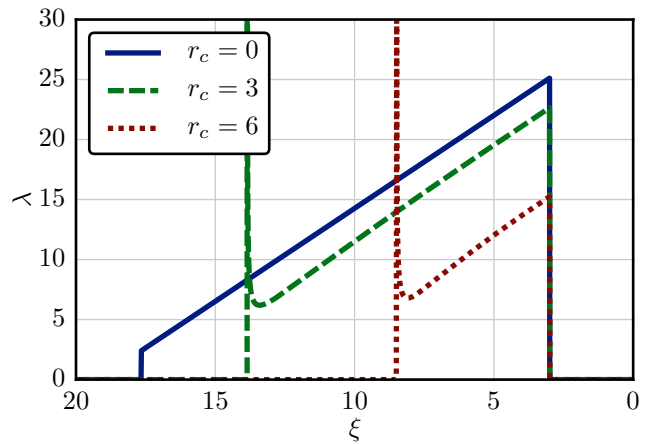}

	\caption{Dependencies of the charge density $\lambda$ in the accelerated bunches on the longitudinal coordinate $\xi$ for different channel radii $r_\chan$.
	The dependencies are calculated numerically using Eqs.~(\ref{constantField:Lambda}) and (\ref{constantField:xi_vs_rb}).
	The maximum bubble radius $R_\subb = 10$, the injection coordinate $\xi_\inj = 3$.}
	\label{fig:constantField:vacuum_profiles}
\end{figure}

Let us consider plasmas with a vacuum channel for which the plasma density is defined as $\rho_\ion(r) = \theta(r_\chan - r)$.
In this case the solutions for the electron bunch profiles providing the homogeneous longitudinal electric field $E_z$ can be found only numerically from Eqs.~(\ref{constantField:Lambda}) and (\ref{constantField:xi_vs_rb}); they are shown in Fig.~\ref{fig:constantField:vacuum_profiles}.
The shapes of these electron bunches are again very close to trapezoidal like in the previous case of the power-law plasma profiles.
They also share the same behavior of having less charge in them for wider channels.

\begin{figure}[tb]
	\includegraphics{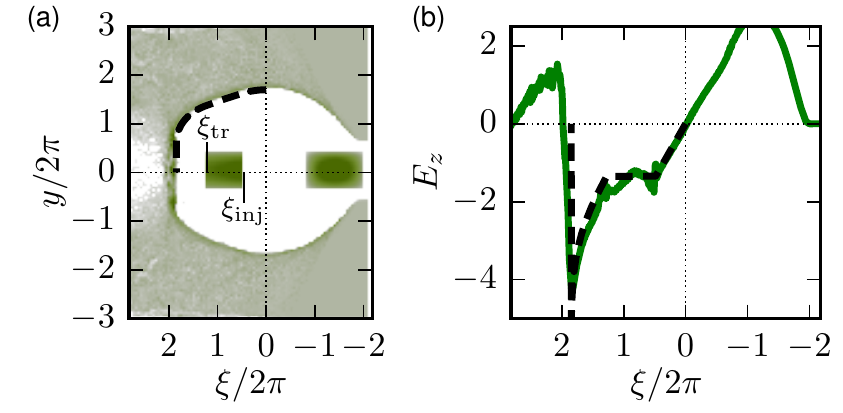}
	\caption{(a)~The bubble and (b)~the time average longitudinal electric field $E_z$ at the axis of the bubble for an electron bunch with the profile chosen according to Eqs.~(\ref{constantField:Lambda}), (\ref{constantField:xi_vs_rb}).
	The channel radius $r_c = 1.2\pi$. 
	Analytical solutions calculated using Eqs.~(\ref{bubbleShape:shapeEquation}) and (\ref{bubbleShape:Ez_initial}) are shown with the dashed lines.
	All lengths are normalized to $\lambda_\plasm = \SI{5}{\um}$.}
	\label{fig:constantField:pic}
\end{figure}

To check if a uniform accelerating field can be generated in plasmas with a channel by using an appropriate electron bunch we carried out 3D PIC simulations.
The results of these simulations are shown in Fig.~\ref{fig:constantField:pic}.
We used plasma with a channel with the density $n_\plasm = \SI{4.5e19}{\cm^{-3}}$ outside the channel.
The bubble in this plasma was driven by an electron bunch which had a total charge of \SI{3.2}{\nano\coulomb} and typical longitudinal and transverse sizes of \SI{6}{\um} and \SI{4}{\um}, respectively.
The accelerated bunch had the same typical width while its longitudinal shape was determined from Eqs.~(\ref{constantField:Lambda}), (\ref{constantField:xi_vs_rb}).
The energy of the particles in both bunches was \SI{30}{\GeV}.
The comparison of the results of these simulations to the analytical solutions in the scope of our model shows good correspondence between them.

\section{Conclusions}
The equation for the bubble boundary in transversely inhomogeneous plasmas derived in \citet{Thomas2015channels} is analytically studied.
The boundary of a non-loaded bubble is found for an arbitrary plasma profile.
The efficiency of the energy transfer from the bubble to an arbitrary accelerated electron bunch is described.
When the accelerated bunch is flat-top the threshold value of the bunch charge density which provides maximum efficiency of the acceleration is found.
It is also shown that the artificial selection of the electron bunch profile can be used to create homogeneous longitudinal field in the volume of the bunch, which is important for increasing the quality of the accelerated bunch.

The general results are applied to plasmas with power-law profiles and with vacuum channels.
It is shown that the increase of the channel size or its depth leads to the contraction of the bubble and to the decrease of the electric field gradient in it.
For these plasma profiles the possibility of creating homogeneous accelerating field by the adjustment of the density profile of the electron bunch is demonstrated.
The shape of these bunch profiles is proved to be close to trapezoidal.

For the case of a vacuum channel 3D PIC simulations are carried out.
They demonstrate good correspondence with the predictions of the model both for non-loaded bubbles and for a bubble with the accelerated bunch providing homogeneous electric field.

\begin{acknowledgments}
This work has been supported by the Russian Science Foundation through Grant No. 16-12-10383.
\end{acknowledgments}

\section*{References}
\bibliographystyle{aipnum4-1}
\bibliography{BibliographyPop}

\end{document}